\documentclass[conference]{IEEEtran}
\IEEEoverridecommandlockouts

\usepackage{color}
\usepackage{cite}
\usepackage{tabularx}
\usepackage{tikz}
\usepackage{soul}
\usepackage{multirow}
\usepackage{cite}
 
\usepackage{graphicx}
\usepackage{float}
\usepackage{subfigure}
\usepackage{cite} 
\usepackage{amsmath}

\usepackage{cite}
\usepackage{amsmath,amsfonts}
\usepackage{algorithm}
\usepackage{array}
\usepackage{algpseudocode}
\usepackage[caption=false,font=normalsize,labelfont=sf,textfont=sf]{subfig}
\usepackage{textcomp}
\usepackage{stfloats}
\usepackage{url}
\usepackage{verbatim}
\usepackage{graphicx}
\usepackage{subfigure}
\usepackage{pifont}
\usepackage{makecell}
\usepackage{bm}
\usepackage{bbm}
\usepackage[english]{babel}
\usepackage{amsthm}
\usepackage{subcaption}
\usepackage{color}
\usepackage{subfigure}

\IEEEoverridecommandlockouts

\usepackage{tabularx}
\usepackage{tikz}
\usepackage{soul}
\usepackage{multirow}
 
\usepackage{graphicx}
\usepackage{float}
\begin{document}

\title{Gesture-Aware Indoor THz ISAC Systems for Adaptive Resource Allocation}

\author{\IEEEauthorblockN{Zhonghao Liu\IEEEauthorrefmark{1}, 
Yinchao Yang\IEEEauthorrefmark{1}, Yahao Ding\IEEEauthorrefmark{1}, Yixuan Wang\IEEEauthorrefmark{1}, and Mohammad Shikh-Bahaei\IEEEauthorrefmark{1}}
\IEEEauthorblockA{\IEEEauthorrefmark{1}King's College London\\
Email: \{zhonghao.liu, yinchao.yang, yahao.ding, k21040372, m.sbahaei\}@kcl.ac.uk}}

% The paper headers
\markboth{Journal of \LaTeX\ Class Files,~Vol.~14, No.~8, August~2015}%
{Shell \MakeLowercase{\textit{et al.}}: Bare Demo of IEEEtran.cls for IEEE Journals}

\maketitle

\begin{abstract}
%This paper investigates a multi-user indoor integrated sensing and communication (ISAC) system operating in the terahertz (THz) band, designed for gesture recognition in smart room. With a tracking model, the system adaptively adjusts resource allocation in response to gesture changes to enhance overall sensing performance. To address the varying communication quality requirements reflected by different user gestures, a adaptive joint optimization algorithm for power allocation and beamforming is proposed, aiming to maximize the total sensing signal-to-interference-plus-noise ratio (SINR) while satisfying the quality of communication. Simulation results demonstrate that the proposed method effectively reacts to gesture variations and significantly improves sensing performance, outperforming single-variable optimization baselines.
This paper investigates a multi-user indoor integrated sensing and communication (ISAC) system operating in the terahertz (THz) band, designed for adaptive communication based on gesture recognition. Leveraging gesture tracking through an extended Kalman filter (EKF), the access point (AP) dynamically adjusts resource allocation in response to detected gesture variations, thereby improving sensing accuracy. Based on the gesture recognition results, the AP further updates the communication quality requirements of different users, enabling efficient resource allocation. To this end, an adaptive joint optimization algorithm for power allocation and beamforming is developed to maximize the overall sensing signal-to-interference-plus-noise ratio (SINR) while satisfying the gesture-dependent communication quality of service (QoS) constraints. Simulation results demonstrate that the proposed method effectively responds to gesture dynamics, achieving superior sensing accuracy and communication performance compared with conventional single-variable optimization baselines.
\end{abstract}


\section{Introduction}
 %As the sixth generation (6G) of mobile communications approaches, wireless networks are evolving from traditional data transmission platforms into intelligent systems capable of environmental sensing, autonomous decision-making, and adaptive control. Integrated sensing and communication (ISAC), as a key enabler of this transformation, aims to perform both communication and sensing tasks within a unified system architecture and spectrum resource, thereby significantly improving spectral efficiency and reducing system overhead. ISAC is empowering a wide range of intelligent applications such as autonomous driving, smart home, and human–machine interaction \cite{10293761}. The terahertz(THz) band offers inherently finer sensing resolution due to its shorter wavelength and supports higher communication rates enabled by its wide contiguous bandwidth~\cite{shang2024unlocking}. Accordingly, THz-based ISAC is well-suited to compact and dynamic environments, particularly indoor gesture recognition.
 As 6G approaches, wireless networks are evolving from conventional data transmission platforms into intelligent systems capable of environmental sensing, autonomous decision-making, and adaptive control. Integrated sensing and communication (ISAC), a key enabler of this paradigm shift, aims to jointly perform communication and sensing functions within a unified system architecture and shared spectral resources, thereby enhancing spectral efficiency and reducing system overhead. ISAC technology is paving the way for a wide range of intelligent applications, including autonomous driving, smart homes, and human machine interaction~\cite{10293761}. In immersive XR/VR and smart indoor spaces, human gestures act as a natural control interface and often coincide with abrupt changes in traffic demand and latency constraints, which calls for sensing-assisted and context-aware resource adaptation.
 The terahertz (THz) band provides an attractive platform for ISAC systems because its short wavelength enables high resolution sensing, and its wide bandwidth supports high data rates~\cite{shang2024unlocking}. Consequently, THz-based ISAC is particularly well-suited to compact and dynamic indoor environments, such as gesture recognition scenarios.
 
 In indoor environments, communication requirements are highly dynamic, as user demands vary with their behavioral state. For example, picking up a smart device typically indicates an active interaction session (e.g., video call, messaging, XR control), whereas putting it down often corresponds to an idle state. As a result, the QoS constraints may change abruptly over time, making static designs ineffective. Consequently, intelligent and adaptive resource allocation is required to track user behavior and jointly accommodate time-varying communication demands and sensing tasks.
 
 %In previous studies, extensive efforts have been devoted to THz or ISAC systems. For instance, Liu \textit{et al.}~\cite{li2024enhancing} predicted user trajectories and velocities based on sensing data to enable timely beam adjustment and avoid communication blockages. Fang \textit{et al.}~\cite{fang2025environment} achieved fine-grained environmental reconstruction and material identification, verifying the high resolution and feasibility of THz sensing in complex indoor scenarios. Zhou \textit{et al.}~\cite{10681603} improved sensing accuracy through optimized beamforming and transmission design while maintaining communication quality. Liu \textit{et al.}~\cite{10439221} systematically reviewed the key challenges and solutions of ISAC in the THz band. In addition, adaptive communication techniques under dynamic environments have also been studied. For example, Zhou \textit{et al.}~\cite{bacchiellibistatic} proposed a THz MIMO-OFDM ISAC framework for cluttered indoor environments, which enables target tracking and other sensing tasks while maintaining high-rate communication. However, these papers fail to consider dynamic variations and their impact on real-time resource optimization in THz-ISAC indoor systems. 

 In previous studies, extensive research efforts have been devoted to THz and ISAC systems. For instance, Li \textit{et al.}~\cite{10438962} leveraged sensing data to predict user trajectories and velocities, thereby enabling timely beam adjustments and mitigating communication blockages. Fang \textit{et al.}~\cite{fang2025environment} demonstrated fine-grained environmental reconstruction and material identification, verifying the high resolution and feasibility of THz sensing in complex indoor environments. Zhou \textit{et al.}~\cite{zhou2024near} enhanced sensing accuracy through optimized beamforming and transmission design, while maintaining communication quality. Furthermore, adaptive communication techniques for dynamic environments have also been explored. For example, Bacchielli \textit{et al.}~\cite{bacchiellibistatic} proposed a THz MIMO-OFDM ISAC framework for cluttered indoor environments, enabling target tracking and other sensing functions while sustaining high data rates. However, most existing works primarily focus on static scenarios and overlook the dynamic variations in user gestures and behaviors, as well as their impact on real-time resource optimization in THz-ISAC indoor systems.

Inspired by these works, we propose a sensing-assisted THz ISAC system that dynamically infers the user's communication requirements based on the recognized gesture results, enabling more efficient resource allocation between sensing and communication. The system continuously monitors hand gestures (e.g., picking up or putting down a device) to infer the user activity state and adjust the QoS accordingly. When active usage is detected, more power is allocated to communication; otherwise, communication power is reduced to suppress interference and improve sensing performance. 

The main contributions of this work are summarized as follows:
\begin{itemize}
   \item We investigate an indoor THz-ISAC system under dynamic user scenarios, where gesture variations lead to time-varying communication and sensing demands.

  \item We construct a mapping mechanism from gesture states to communication QoS requirements, through which the user’s communication QoS demand can be inferred from the recognized gesture state.
  
  \item We design a tracking module based on echo signal processing that predicts gesture transitions and provides real-time state estimation for adaptive resource management.
\end{itemize}

\addtolength{\topmargin}{0.06in}
\section{system model}
\enlargethispage{-0.10in}
\begin{figure}[!t]
    \centering
    \includegraphics[width=0.5\textwidth]{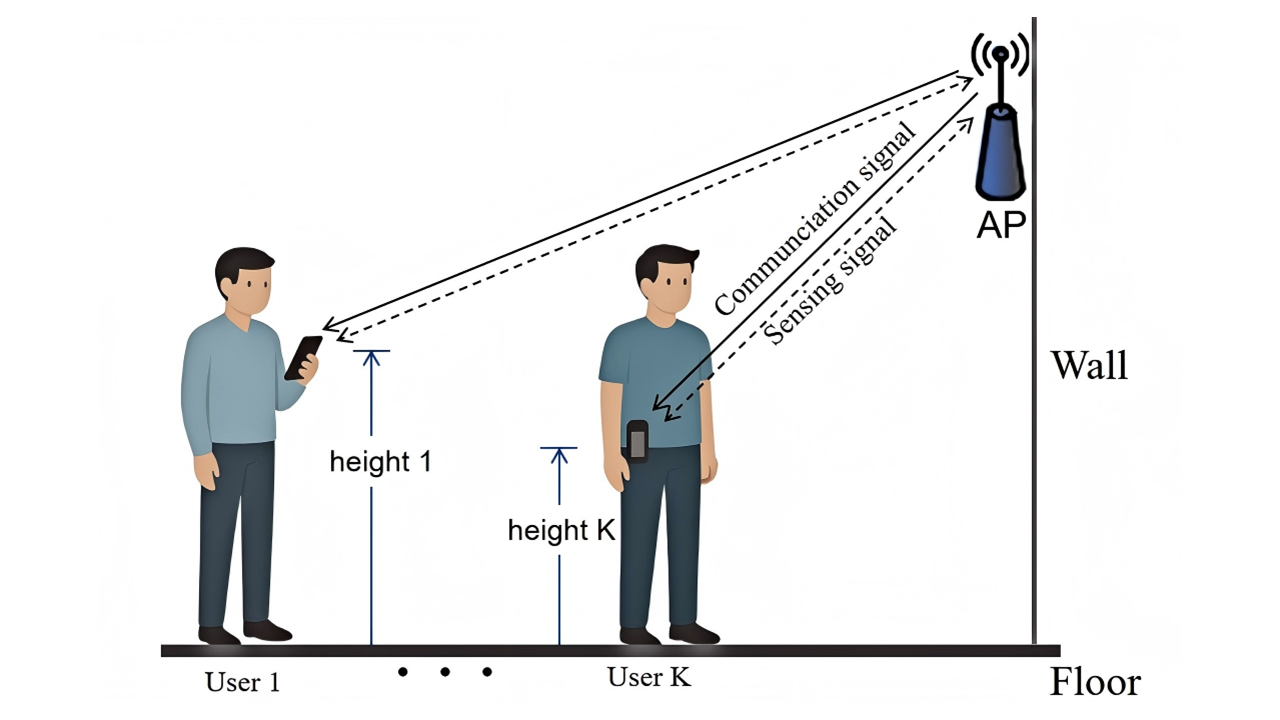}  % 替换为你自己的图片文件名
    \caption{The indoor THz-ISAC system, where an access point is deployed at the corner of the room and transmits both communication and sensing signals.}
    \label{fig:1}
\end{figure}

As shown in Fig.~\ref{fig:1}, we consider a THz-ISAC system, where an access point (AP) simultaneously serves $K$ single-antenna users and performs gesture sensing. The AP is equipped with a uniform linear array (ULA) consisting of $M$ antennas. Specifically, hand gestures such as picking up or putting down a device are modeled as changes in the vertical position of the user's hand. For simplicity, the device height is assumed to be $h_A$ after picking up and $h_B$ after putting down.

\subsection{Signal Model}
\enlargethispage{-0.10in}
At time slot $l$, the AP transmits a dual-functional signal to support both communication and sensing, which can be expressed as:
\begin{equation}
    \mathbf{x}[l] = \mathbf{W}_c  \mathbf{s}_c[l] + \mathbf{W}_rs_r[l],
\end{equation}
where $\mathbf{W}_c=[\sqrt{P_1}\mathbf{w}_{c,1},\ldots,\sqrt{P_K}\mathbf{w}_{c,K}] \in \mathbb{C}^{M\times K}$
denotes the communication beamforming matrix, and
$\mathbf{W}_r = \sqrt{P_r}\,[\mathbf{w}_{r,1},\ldots,\mathbf{w}_{r,K}] \in \mathbb{C}^{M\times K}$
denotes the sensing beamforming matrix. Here, $P_k$ denotes the transmit power allocated to the user $k$ and the sensing power $P_r$ is uniformly applied across all sensing beams.  Each user~$k$ is associated with a communication beamforming vector~$\mathbf{w}_{c,k} \in \mathbb{C}^{M \times 1}$ and a sensing beamforming vector~$\mathbf{w}_{r,k} \in \mathbb{C}^{M \times 1}$.
The communication symbols for the $K$ users at time slot $l$ are denoted by $\mathbf{s}_c[l] \in \mathbb{C}^{K}$, which consists of mutually orthogonal symbols across different users, such that $\mathbb{E}[\mathbf{s}_c[l]\mathbf{s}_c^H[l]] = \mathbf{I}_K$. Similarly, the radar signal vector contains $K$ orthogonal waveforms for $K$ users, denoted by $\mathbf{s}_r[l] \in \mathbb{C}^{K}$, satisfying $\mathbb{E}[\mathbf{s}_r[l]\mathbf{s}_r^H[l]] = \mathbf{I}_K$. Furthermore, communication symbols and radar waveforms are assumed to be statistically independent, satisfying $\mathbb{E}[\mathbf{s}_c[l]\mathbf{s}_r^H[l]] = 0$.

In the proposed ISAC system, the AP transmits the dual-functional signal to the users. Therefore, the received signal by user $k$ at time slot $l$ is given by~\cite{towhidlou2018adaptive}:
\begin{align}
y_k[l] &= \mathbf{h}_k^H \sqrt{P_k} \mathbf{w}_{c,k} s_{c,k}[l] 
+ \sum_{j \neq k} \mathbf{h}_k^H \sqrt{P_j} \mathbf{w}_{c,j} s_{c,j}[l] \notag \\
&\quad + \mathbf{h}_k^H \sqrt{P_r} \mathbf{w}_{r,k} s_{r,k}[l] 
+ n_k[l],
\end{align}
where \( \mathbf{h}_k \in \mathbb{C}^{M \times 1} \)is the wireless channel vector of the user $k$, the detailed modeling of $\mathbf{h}_k$ will be introduced in the next subsection, and $n_k[l] \sim \mathcal{CN}(0, \sigma_n^2 )$ is the Gaussian noise.
The received echo signal can be expressed as:
\begin{equation}
\label{G}
     \mathbf{y}_{\text{echo}}[l] =  \sum_{k=1}^{K} \mathbf{G}_k^H \mathbf{x}[l - \tau_k] + \mathbf{n}_r[l],
\end{equation}
where \( \mathbf{G}_k \in \mathbb{C}^{M \times M} \) is the channel matrix of user $k$, and its formulation will be discussed in the next subsection. In addition, \( \tau_k \) denotes the propagation delay of the echo signal and $\mathbf{n}_r[l] \sim \mathcal{CN}(0, \sigma_r^2 )$ is the Gaussian noise.
\subsection{Channel Model}
\enlargethispage{-0.10in}
Due to the significant path loss and inherent sparsity of THz waveforms \cite{10223713}, we focus solely on the line-of-sight (LoS) channel. The channel between the AP and user $k$ is given by:  
\begin{equation}
    \mathbf{h}_k =  \frac{c}{4\pi f d_k} e^{-\frac{1}{2}K(f)d_k} \mathbf{a}(\theta_k),
\end{equation}
where the term $ \frac{c}{4\pi f d_k} e^{-\frac{1}{2}K(f)d_k}$ represents the THz path loss. Here, $c$ is the speed of light, $f$ is the carrier frequency, $d_k$ is the distance between the AP and user $k$, and $K(f)$ is the molecular absorption coefficient at frequency $f$, which can be obtained from HITRAN database\cite{9667397}. The term $\mathbf{a}(\theta_k)$ denotes the array response vector at the AP for the departure angle $\theta_k$, given as $\mathbf{a}(\theta_k) = [1, e^{j \frac{2\pi d}{\lambda} \sin \theta_k}, ..., e^{j \frac{2\pi (M-1) d}{\lambda} \sin \theta_k}]^T$.

Similarly, the target’s reflection channel can be modeled as:  
\begin{equation}
     \mathbf{G}_k = \left(\frac{c}{4\pi f d_k}\right)^{2}
e^{-K(f)\, d_k}\,
\beta_k \,
e^{\,-j \tfrac{4\pi f v_{k} t}{c} }\mathbf a(\theta_k)\mathbf a^{H}(\theta_k),
\end{equation}
where $\beta_k$ is the radar cross-section (RCS) of the user $k$.

\subsection{Performance Measurement}
\enlargethispage{-0.10in}
We adopt signal-to-interference-plus-noise ratio (SINR) as the performance metric for both communication and sensing. The communication SINR of user $k$ at time slot $l$ is expressed as: 
\begin{equation}
\label{eq:adaptive_com_sinr}
\begin{aligned}
&\text{SINR}_{com,l}^{k} =\\ 
&\frac{P_{k,l} |\mathbf{h}_{k,l}^H \mathbf{w}_{c,k,l}|^2}
{\sum_{j=1, j \neq k}^K P_{j,l} |\mathbf{h}_{k,l}^H \mathbf{w}_{c,j,l}|^2 
+ P_{r,l} \sum_{j=1}^{K} |\mathbf{h}_{k,l}^H \mathbf{w}_{r,j,l}|^2 
+ \sigma_n^2},
\end{aligned}
\end{equation}
where the subscript $l$ indexes the time slot $l$.

When the user picks up the device, for example, to send a message, it indicates a higher communication demand, and thus a higher communication SINR is required, denoted by \( \mathrm{SINR}_{\mathrm{com}}^{\text{high}} \). Conversely, when the user puts down the device, the communication demand is reduced, and a lower SINR 
\addtolength{\topmargin}{0.06in}
requirement \( \mathrm{SINR}_{\mathrm{com}}^{\text{low}} \) is applied.
We define the required communication SINR threshold of user $k$ as a function of gesture state, expressed as: 
\begin{equation}
\label{7}
\gamma_{k,l}^{\rm req}
 = \delta_{k,l} \mathrm{SINR}_{com}^{\text{high}} + (1 - \delta_{k,l})  \mathrm{SINR}_{com}^{\text{low}}, 
\end{equation}
where $\delta_{k,l} \in \{0, 1\}$ is a binary indicator representing the communication requirement of user $k$ at time slot $l$, which is dependent on the gesture. When the user is recognized as picking up the device, we set $\delta_{k,l} = 1$ to impose a higher communication requirement; otherwise, we set $\delta_{k,l} = 0$. 

\enlargethispage{-0.10in}
In the modeling of the reflected sensing signals, we assume that the number of antennas at the AP satisfies \( M \gg K \). Under this condition, the reflected signals from different users exhibit strong spatial separability, and the mutual interference between users becomes negligible~\cite{liu2023snr}. Therefore, the sensing SINR is given by:
\begin{equation}
\label{eq:adaptive_sen_sinr}
\text{SINR}_{sen,l}^{k} = 
\frac{P_{r,l} |\mathbf{G}_{k,l} \mathbf{w}_{r,k,l}|^2}
{P_{k,l} |\mathbf{G}_{k,l} \mathbf{w}_{c,k,l}|^2 + \sigma_r^2}.
\end{equation}
%where all symbols follow the previous definitions and the subscript $l$ indexes the $l$-th time slot. 

\section{Gesture movement model and State prediction}
\subsection{Gesture state movement model}
\enlargethispage{-0.10in}
The gesture motion is modeled as a discrete-time state-space system under the assumption of constant velocity within each time slot \( l \). %We assume that the echo signal of user \( k \) at time slot \( l \) is obtained at the AP via a filtering technique. The channel state of the user at time slot \( l+1 \) is then predicted by analyzing the echo information at slot \( l \). Subsequently, the filter is updated using the echo received at slot \( l+1 \) to reduce the prediction error for the next time slot. 
%At the \( l \)-th time slot, $\tau_{l,k}$ can be obtained from the received pilot signals. Based on that, the distance between the antenna and the hand of the \( k \)-th user can be expressed as \( d_{l,k} = \frac{c \cdot \tau_{l,k}}{2} \).
Using a matched filter, the time delay $\tau_{l,k}$ and the Doppler shift $f_l$ can be estimated. Based on that, the distance between the antenna and the hand of the user \( k \)  can be expressed as \( d_{l,k} = \frac{c \cdot \tau_{l,k}}{2} \) with $c$ being the speed of light.

The constant speed \( v_{l,k} \) can be decomposed into a radial component \( v_{l,k}^r \) and a tangential component \( v_{l,k}^t \) at the time slot \( l \), with respect to the line connecting the hand and the antenna. We assume that $\theta_{l,k}$ denotes the angle of arrival (AoA) of user $k$ at time slot $l$. The radial and tangential velocities $v_{l,k}^r$ and $v_{l,k}^t$ are treated as known inputs which can be obtained following the model in \cite{zhang2023robust}. Therefore, the distance and AoA for user \( k \) at the next time slot are given by:
\begin{equation}\label{eq4}
d_{l,k} = d_{l-1,k} - v_{l-1,k}^r T, \quad \theta_{l,k} = \theta_{l-1,k} + \frac{v_{l-1,k}^t T}{d_{l-1,k}},
\end{equation}
where \( T \) is the length of the time slot.

%We derive the transformation from a polar-coordinate-based motion model to Cartesian velocity components. Given the  \( \theta_{l,k} \) at time slot \( l \), the radial and tangential velocities can be expressed in terms of the Cartesian velocity components \( v_{l,k}^x \) and \( v_{l,k}^y \) as follows\cite{zhang2023robust}:
%\begin{equation}
%\left\{
%\begin{aligned}
%v_{l,k}^r &= v_{l,k}^x \cos \theta_{l,k} + v_{l,k}^y %\sin \theta_{l,k}, \\
%v_{l,k}^t &= -v_{l,k}^x \sin \theta_{l,k} + v_{l,k}^y \cos \theta_{l,k},
%\end{aligned}
%\right.
%\end{equation}
%where \( v_{l,k}^x \) and \( v_{l,k}^y \) are estimated by differencing the positions obtained from two consecutive echo measurements, as \( v_{l,k}^x = \frac{x_{l,k} - x_{l-1,k}}{T} \) and \( v_{l,k}^y = \frac{y_{l,k} - y_{l-1,k}}{T} \).

We observe that the device height typically increases when the users pick up the device and decreases when the device is put down. Therefore, we use the variation in device height as the primary feature for gesture recognition. Specifically, the height change is defined as \( \Delta h_{l,k} = h_{l,k} - h_{l-1,k} \), where the device height at time slot \( l \) is calculated by \( h_{l,k} = d_{l,k} \cos\theta_{l,k} \). Therefore, based on the echo information obtained at time slot $l$ and the predicted distance and angle at time slot $l+1$, we compute the height variation to determine whether a gesture has occurred. To determine whether a gesture actually occurs (as opposed to small fluctuations or noise), we introduce a height variation threshold \( \epsilon_h \). 
Therefore, the decision rule for gesture state change is defined as:
\begin{equation}
\label{eq:gesture_rule}
\begin{cases}
\text{inactive}, & |\Delta h_{l,k}| < \epsilon_h,\\[2pt]
\text{picking up},       & |\Delta h_{l,k}| \ge \epsilon_h \text{ and } \Delta h_{l,k} > 0,\\[2pt]
\text{putting down},     & |\Delta h_{l,k}| \ge \epsilon_h \text{ and } \Delta h_{l,k} < 0.
\end{cases}
\end{equation}

Once the system recognizes that user $k$ is picking up the device, the communication SINR is immediately raised to  \( \mathrm{SINR}_{\mathrm{com}}^{\text{high}} \), $\delta_{k,l}$ is set to $1$. Conversely, when the device is put down, $\delta_{k,l}$ is set to $0$. If the gesture state is inactive, $\delta_{k,l}$ retains its previous value.

\subsection{Extended Kalman Filtering}
\enlargethispage{-0.20in}
In this part, we propose an extended Kalman filter (EKF) for gesture state prediction and tracking. In the system with $K$ users, the state of the user $k$ at time slot $l-1$ is defined by the distance and the AoA relative to the AP, and the state vector is denoted as $\mathbf{x}_{l-1,k} = [d_{l-1,k},\ \theta_{l-1,k}]^\top$. The measured vector at time slot $l$ is denoted as $\mathbf{z}_{l,k} = [\tau_{l,k},\ \theta_{l,k}]^\top$. Thus, the model can be recast in compact forms as:
\begin{equation}
\left\{
\begin{aligned}
\text{State Evolution Model: } &\quad \mathbf{x}_{l,k} = \mathbf{f}(\mathbf{x}_{l-1,k}) + \mathbf{Q}_{l,k},\\
\text{Measurement Model: } &\quad \mathbf{z}_{l,k} = \mathbf{h}(\mathbf{x}_{l,k}) + \mathbf{R}_{l,k},
\end{aligned}
\right.
\end{equation}
where $\mathbf{f}(\cdot)$ is defined in~\eqref{eq4}, and 
$\mathbf{Q}_{l,k} = [Q_{\theta,k}, Q_{d,k}]^T$ denotes the process noise vector, which is independent of $\mathbf{f}(\mathbf{x}_{l-1,k})$. According to $\mathbf{y}_{\text{echo}}$ in ~\eqref{G}, we can obtain measurement model for user $k$. 
%Thus, $\mathbf{h}(\cdot)$ is defined as $h(\mathbf{x}_{l,k}) = [\frac{2 d_{l,k}}{c},\ \theta_{l,k}]^T$. 
Similarly, the measurement noise $\mathbf{R}_{l,k}=[R_{\tau,k}, R_{\theta,k}]^T$ is independent of $\mathbf{h}(\mathbf{x}_{l-1,k})$. As considered above, both $\mathbf{Q}_{l,k}$ and $\mathbf{R}_{l,k}$ follow zero-mean Gaussian distributions, with covariance matrices being expressed as:
\begin{align}
\mathbf{Q}_{s} = \mathrm{diag}(\sigma_{d}^{2},~\sigma_{\theta}^{2}), \mathbf{R}_{m} = \mathrm{diag}(\sigma_{\tau}^{2},~\sigma_{\theta}^{2}).
\end{align}
%The temporal evolution of the state follows \eqref{eq4}, where the state of the $k$-th user at time slot $l$ is given by $\mathbf{x}_{l,k} = f^k(\mathbf{x}_{l-1,k}) = [d_{l-1,k} - v_{l-1,k}^r T,\ \theta_{l-1,k} + \frac{v_{l-1,k}^t T}{d_{l-1,k}}]^\top$. On the observation side, the system measures TOA and AOA. Since TOA is linearly related to the distance as $\tau_{l,k} = \frac{2 d_{l,k}}{c}$, the observation vector is defined as $\mathbf{z}_{l,k} = h^k(\mathbf{x}_{l-1,k}) = [\frac{2 d_{l-1,k}}{c},\ \theta_{l-1,k}]^\top$.
In order to linearize the models, the Jacobian matrices for both $\mathbf{f}(\mathbf{x}_{l,k})$ and $\mathbf{h}(\mathbf{x}_{l,k})$
need to be computed.
By straightforward differentiation, the Jacobian of $\mathbf{f}(\mathbf{x}_{l-1,k})$ is given as:
\begin{equation}
\label{eq:f}
\left.\frac{\partial \mathbf{f}}{\partial \mathbf{x}}\right|_{\mathbf{x} = \hat{\mathbf{x}}_{l-1,k}} =
\begin{bmatrix}
1 & 0 \\
- \dfrac{v_{l-1,k}^t T}{(d_{l-1,k})^2} & 1
\end{bmatrix}.
\end{equation}
 The Jacobian of $\mathbf{h}(\mathbf{x}_{l,k})$ is:
\begin{equation}
\label{eq:h}
 \left.\frac{\partial \mathbf{h}}{\partial \mathbf{x}}\right|_{\mathbf{x} = \hat{\mathbf{x}}_{l,k}} =
\begin{bmatrix}
\dfrac{2}{c} & 0 \\
0 & 1
\end{bmatrix}.
\end{equation}
Based on ~\eqref{eq:f} and ~\eqref{eq:h}, following the standard procedure of Kalman filtering~\cite{Simon2006}, the gesture state prediction and tracking design is summarized as follows:
\subsubsection{State Prediction}
\begin{equation}
\hat{\mathbf{x}}_{l,k|l-1,k} = \mathbf{f}(\hat{\mathbf{x}}_{l-1,k}).
\label{eq:state-prediction}
\end{equation}
\subsubsection{Linearization}
\begin{equation}
\mathbf{F}_{l-1,k} = 
\left.\frac{\partial \mathbf{f}}{\partial \mathbf{x}}\right|_{\mathbf{x} = \hat{\mathbf{x}}_{l-1,k}},
\mathbf{H}_{l,k} =
\left.\frac{\partial \mathbf{h}}{\partial \mathbf{x}}\right|_{\mathbf{x} = \hat{\mathbf{x}}_{l,k|l-1,k}}.
\label{eq:linearization}
\end{equation}
\subsubsection{MSE Matrix Prediction}
\begin{equation}
\mathbf{M}_{l,k|l-1,k} = 
\mathbf{F}_{l-1,k}\mathbf{M}_{l-1,k}\mathbf{F}_{l-1,k}^{H} + \mathbf{Q}_{s}.
\label{eq:mse-prediction}
\end{equation}
\addtolength{\topmargin}{0.06in}
\subsubsection{Kalman Gain Calculation}
\begin{equation}
\mathbf{G}_{l,k} = 
\mathbf{M}_{l,k|l-1,k}\mathbf{H}_{l,k}^{H}
\left(
\mathbf{R}_{m} + 
\mathbf{H}_{l,k}\mathbf{M}_{l,k|l-1,k}\mathbf{H}_{l,k}^{H}
\right)^{-1}.
\label{eq:kalman-gain}
\end{equation}
\subsubsection{State Tracking}
\begin{equation}
\hat{\mathbf{x}}_{l,k} = 
\hat{\mathbf{x}}_{l,k|l-1,k} + 
\mathbf{G}_{l,k}
\left(
\mathbf{z}_{l,k} - 
\mathbf{h}\!\left(\hat{\mathbf{x}}_{l,k|l-1,k}\right)
\right).
\label{eq:state-tracking}
\end{equation}
\subsubsection{MSE Matrix Update}
\begin{equation}
\mathbf{M}_{l,k} = 
\left(\mathbf{I} - \mathbf{G}_{l,k}\mathbf{H}_{l,k}\right)
\mathbf{M}_{l,k|l-1,k}.
\label{eq:mse-update}
\end{equation}
%We first process the echoes from slot~$l-1$
%to obtain the posterior state estimate $\mathbf{x}_{l-1|l-1,k}$ and state estimation error covariance $\mathbf{P}_{l-1|l-1,k}$for user~$k$. Based on that,
%the state transition model yields the prior estimate $\hat{\mathbf{x}}_{l|l-1,k}$ and estimate error covariance $\hat{\mathbf{P}}_{l|l-1,k}$ for the time slot $l$, which can be donate as:
%\begin{align}
%\label{14}
       % \hat{\mathbf{x}}_{l|l-%1,k} &= %f^k(\hat{\mathbf{x}}_{l-%1|l-1,k}), \\
       % \hat{\mathbf{P}}_{l|l-%1,k} &= \mathbf{F}_{l-%1,k} \mathbf{P}_{l-1|l-%1,k} \mathbf{F}_{l-%1,k}^\top + \mathbf{Q}_{l-%1,k},
       % \label{15}
   % \end{align}
%where \( \mathbf{Q}_{l-1,k} \) denotes the process noise covariance matrix.
%\textbf{Step 2 (Update)}
%After receiving the echo signal of time slot $l$, the update step is performed. The Kalman gain is given by
%\begin{align}
%\label{16}
  %  \mathbf{K}_{l,k} &= \mathbf{P}_{l|l-1,k} \mathbf{H}_{l,k}^\top \left( \mathbf{H}_{l,k} \mathbf{P}_{l|l-1,k} \mathbf{H}_{l,k}^\top + \mathbf{R}_{l,k} \right)^{-1}, 
%\end{align}
%where \( \mathbf{R}_{l,k} \) is the observation noise covariance. Then, the posterior state estimate $\mathbf{x}_{l|l,k}$ and state estimation error covariance $\mathbf{P}_{l|l,k}$ of time slot $l$ are updated as
%\begin{align}
%\label{17}
%\mathbf{x}_{l|l,k} &= \hat{\mathbf{x}}_{l|l-1,k}
%+ %\mathbf{K}_{l,k}\big(\mathbf{z}_{l,k}-h^k(\hat{\mathbf{x}}_{l|l-1,k})\big), \\
%\mathbf{P}_{l|l,k} &= \big(\mathbf{I}-\mathbf{K}_{l,k}\mathbf{H}_{l,k}\big)
%\,\hat{\mathbf{P}}_{l|l-1,k}.\label{18}
%\end{align}

\section{Problem Formulation}
\enlargethispage{-0.30in}
During time slot $l$, we aim to jointly optimize the beamforming vector and power to maximize the sum sensing SINR while ensuring the power constraints and communication SINR requirements. We denote the optimization variables at time slot $l$ as 
$P_{r,l}, P_{k,l}, \mathbf{w}_{c,k,l}, \mathbf{w}_{r,k,l}$, 
and the sensing channel as $\mathbf{G}_{k,l}$. The optimization problem shown as:
\begin{subequations}\label{opt1}
\begin{align}
   &\max_{P_{k,l}\, P_{r,l}, \mathbf{w}_{c,k,l}, \mathbf{w}_{r,k,l}}\quad \label{opt1a}
 \sum_{k=1}^{K} \frac{P_{r,l} | \mathbf{G}_{k,l} \mathbf{w}_{r,k,l} |^2}
{P_{k,l} | \mathbf{G}_{k,l} \mathbf{w}_{c,k,l} |^2 + \sigma_r^2} \\
    & \text{s.t.} \quad
    KP_{r,l} + \sum_{k=1}^{K} P_{k,l} \leq P_{\max},  \label{opt1b}\\
    & \quad\quad P_{r,l} \geq 0, \quad P_{k,l} \geq 0, \quad \forall k,  \label{opt1c}\\
    & \quad\quad \|\mathbf{w}_{c,k,l}\|^2 = 1, \forall k,  \label{opt1d}\\
    & \quad\quad \|\mathbf{w}_{r,k,l}\|^2 = 1, \forall k,  \label{opt1e}\\
    & \quad\quad \mathrm{SINR}_\text{com,l}^{k} \geq \gamma_{k,l}^{\rm req}, \forall k,  \label{opt1f}
\end{align}
\end{subequations}
where $P_{\text{max}}$ is the total power constraint and \eqref{opt1b} imposes the total transmit-power budget. %In addition, $P_{k,l}$ and $P_{r,l}$ are subject to a total transmit power constraint $P_{\text{max}}$. %both of which are normalized to have unit norm, i.e., $\|\mathbf{w}_{c,k}\| = 1$ and $\|\mathbf{w}_{r,k}\| = 1$. 
Constraints \eqref{opt1d} and \eqref{opt1e} enforce unit–norm constraints on the communication and sensing beamforming vectors $\|\mathbf{w}_{c,k,l}\|$ and $\|\mathbf{w}_{r,k,l}\|$, respectively. 
However, the presence of fractional terms in both the objective function \eqref{opt1a} and the constraints renders the problem non-convex and challenging to solve directly. To address these challenges, we propose utilizing the fractional programming (FP) and alternating optimization (AO) methods to reformulate problem \eqref{opt1} into a tractable convex optimization problem, which is then solved iteratively. 
Since ~\eqref{opt1a} is in the form of a sum of fractional ratios, we apply the quadratic transform and introduce an auxiliary variable \( \mathbf{t_l} = [t_{1,l}, t_{2,l}, \dots, t_{K,l}]^T \), which reformulates it as follows:
{\setlength{\abovedisplayskip}{6pt}
\setlength{\belowdisplayskip}{6pt}
\setlength{\abovedisplayshortskip}{4pt}
\setlength{\belowdisplayshortskip}{4pt}
\begin{align}\label{eq1}
\eqref{opt1a} \rightarrow \sum_{k=1}^{K} 
&2 t_{k,l} \sqrt{P_{r,l} \left| \mathbf{G}_{k,l} \mathbf{w}_{r,k,l} \right|^2} \notag 
\\ 
& - \sum_{k=1}^{K} t_{k,l}^2 \left( P_{k,l} \left| \mathbf{G}_{k,l} \mathbf{w}_{c,k,l} \right|^2 + \sigma_r^2 \right) .
\end{align}

Given other variables, the optimal solution for the auxiliary variable \( t_{k,l} \) can be expressed as follows:
\begin{equation}\label{eq2}
t_{k,l} = 
\frac{
\sqrt{P_{r,l} \left| \mathbf{G}_{k,l} \mathbf{w}_{r,k,l} \right|^2}
}{
P_{k,l} \left| \mathbf{G}_{k,l} \mathbf{w}_{c,k,l} \right|^2 + \sigma_r^2
}.
\end{equation}

Analyzing~\eqref{eq1}, we observe that although \eqref{eq1} is not a convex optimization problem, it becomes convex when optimizing each variable separately while keeping the others fixed. Therefore, we adopt the AO method to transform the original problem into two sub-problems, which can be solved iteratively. By fixing the power variables, we optimize the beamforming vectors \( \mathbf{w}_{c,k,l} \) and \( \mathbf{w}_{r,k,l} \), transforming \eqref{eq1} into the beamforming subproblem. We adopt the semidefinite relaxation (SDR) approach by introducing the matrix variables \( \mathbf{W}_{c,j,l} = \mathbf{w}_{c,j,l} \mathbf{w}_{c,j,l}^H \) and \( \mathbf{W}_{r,j,l} = \mathbf{w}_{r,j,l} \mathbf{w}_{r,j,l}^H \), with the constraints \( \mathbf{W}_{c,j,l} \succeq 0 \) and \( \mathbf{W}_{r,j,l} \succeq 0 \). 
\subsection{Beamforming Optimization}
\enlargethispage{-0.20in}
We formulate the objective function of beamforming subproblem as follows:
{\small
\begin{equation}\label{optf2}
\begin{aligned}
f(\mathbf{W}_{c,k,l}, \mathbf{W}_{r,k,l}, t_{k,l}) &= \sum_{k=1}^{K} \Big( \, 2 t_{k,l} \sqrt{P_{r,l}} \sqrt{\mathrm{Tr}(\mathbf{G}_{k,l} \mathbf{W}_{r,k,l} \mathbf{G}_{k,l}^H)} \\
& - t_{k,l}^2 \left( P_{k,l}\, \mathrm{Tr}(\mathbf{G}_{k,l} \mathbf{W}_{c,k,l} \mathbf{G}_{k,l}^H) + \sigma_r^2 \right) \Big).
\end{aligned}
\end{equation}
}
Meanwhile, ~\eqref{opt1f} is also transformed into a convex constraint, which can be expressed in the following form: 
\begin{equation}\label{eq16}
\begin{aligned}
&P_{k,l}\operatorname{Tr}\!\bigl(\mathbf{W}_{c,k,l}\mathbf{h}_{k,l}\mathbf{h}_{k,l}^H\bigr)
-\gamma_{k,l}^{\rm req}\bigl (
\sum_{j\neq k} P_{j,l}\operatorname{Tr}\!\bigl(\mathbf{W}_{c,j,l}\mathbf{h}_{k,l}\mathbf{h}_{k,l}^H\bigr) \\
&+ P_{r,l}\sum_{j=1}^{K}\operatorname{Tr}\!\bigl(\mathbf{W}_{r,j,l}\mathbf{h}_{k,l}\mathbf{h}_{k,l}^H\bigr)
+\sigma_n^2
\bigr)\geq 0.
\end{aligned}
\end{equation}

Therefore, the corresponding beamforming matrix subproblem is formulated as follows:
\begin{subequations}\label{opt2}
\begin{align}
    & \max_{\mathbf{W}_{c,j,l}, \mathbf{W}_{r,j,l},\; t_{k,l}} 
    && f(\mathbf{W}_{c,j,l}, \mathbf{W}_{r,j,l}, t_{k,l}) \label{opt2a} \\
    & \quad\quad\text{s.t.} 
    && \mathbf{W}_{c,j,l} \succeq 0,\quad \mathbf{W}_{r,j,l} \succeq 0, \quad \forall j \label{opt2b} \\
    & 
    && \text{\eqref{eq16}} \label{opt2c}
\end{align}
\end{subequations}

Problem \eqref{opt2} is transformed into a convex optimization problem and can be efficiently solved using a numerical tool. 

\subsection{Power Allocation}
With the given fixed beamforming matrix \( \mathbf{W}_{c,k,l} \) and \( \mathbf{W}_{r,k,l} \), the original problem reduces to a subproblem with respect to the power variables. The second sub-objective function is defined as:
\begin{equation}\label{optf1}
\begin{aligned}
f(P_{k,l}, P_{r,l}, t_{k,l}) &= \sum_{k=1}^{K} \Big( \, 2 t_{k,l} \sqrt{P_{r,l}} \sqrt{\mathrm{Tr}(\mathbf{G}_{k,l} \mathbf{W}_{r,k,l} \mathbf{G}_{k,l}^H)} \\
& - t_{k,l}^2 \left( P_{k,l}\, \mathrm{Tr}(\mathbf{G}_{k,l} \mathbf{W}_{c,k,l} \mathbf{G}_{k,l}^H) + \sigma_r^2 \right) \Big).
\end{aligned}
\end{equation}
Given this objective, the corresponding power allocation subproblem is formulated as follows:
\begin{subequations}\label{opt3}
\begin{align}
& \max_{P_{k,l},\, P_{r,l},\, t_{k,l}} \quad f(P_{k,l}, P_{r,l}, t_{k,l})  \label{opt3a} \\
& \quad \; \text{s.t.} \quad \eqref{opt1b}, \eqref{opt1c}, \eqref{eq16}. \label{opt3b} 
\end{align}
\end{subequations}

Problem \eqref{opt3} is transformed into a convex optimization problem and can be efficiently solved. Algorithm 1 summarizes the proposed adaptive gesture-aware optimization. The complexity of Algorithm 1 for $L$ timeslot is $\mathcal{O}\!\left(L\left((2KM^2)^{3.5}\log_2(1/\epsilon)+(K+1)^{3.5}\log_2(1/\epsilon)\right)\right)$, where $\epsilon$ is the solution accuracy.

\addtolength{\topmargin}{0.10in}
\begin{algorithm}[t]
\caption{Joint Power and Beamforming Design over Time Slots}
\begin{algorithmic}[1]
\Require $h_{l,k}, G_{l,k}, P_{\max}, \sigma_r^2, \sigma_n^2, \gamma_{k,l}, \forall k, l$
\Ensure $P_{r,l}, P_{k,l}, W_{c,k,l}, W_{r,k,l}, \forall k, l$
\State Initialize $l = 1$ and obtain $\{P_{r,0},P_{k,0},\mathbf W_{c,0},\mathbf W_{r,0}\}$
\While{$l \leq L$ and the users remain within AP coverage} 
    \State Conduct EKF steps given by \eqref{eq:state-prediction}-\eqref{eq:mse-prediction}.
    \State Gesture decision and communication SINR threshold update, as given by \eqref{eq:gesture_rule} and \eqref{7}, respectively. 
    \State Initialize $W_{c,k,l}, W_{r,k,l}$ and $t_{k,l}$.
    \While{not converged}
        \State Update $t_{k,l}, \forall k$, by solving Equation (23).
        \State Update $W_{c,k,l}, W_{r,k,l}$ by solving Equation (26).
        \State Update $t_{k,l}$ by solving Equation (23).
        \State Update $P_{k,l}, P_{r,l}$ by solving Equation (28).
    \EndWhile
    \State Reshape $W_{c,k,l}$ and $W_{r,k,l}$ to obtain rank-one beamforming vectors $w_{c,k,l}$ and $w_{r,k,l}$.
    \State $l \gets l+1$.
    \State Perform the EKF update steps given by \eqref{eq:kalman-gain}-\eqref{eq:mse-update}.
\EndWhile
\State \Return $\{P_{r,l}, P_{k,l}, w_{c,k,l}, w_{r,k,l}\}, \forall k, l$
\end{algorithmic}
\end{algorithm}
\section{Simulation Results}
\enlargethispage{-0.35in}
We conduct numerical simulations to evaluate the proposed model. Following the setting of $K(f)$ in~\cite{9672716}, the simulation parameters are summarized in Table~\ref{table:gesture_sim}. The AP is placed at the upper corner of the room. Each gesture action lasts about 1 second. The simulation proceeds in discrete time slots of 0.1 seconds, corresponding to 10 time slots per complete gesture.

\subsection{Performance under Static Gesture States}
We first consider a static scenario with fixed user positions and gesture states, where the communication QoS are available as prior information, and evaluate performance under different antenna numbers and power budgets.

We compare the proposed design with power-only \cite{10557534} and optimize beam only \cite{10086626}. As shown in Fig.~\ref{fig2}, when the total transmission power constraint increases, both the joint optimization and power-only schemes show slight improvements in sensing performance. This is because, under proportional power initialization and beam-only optimization, the increased communication leakage outweighs the sensing gain. As shown in Fig.~\ref{fig3}, the sensing SINR increases with the number of antennas, since a larger array provides higher array gain and finer angular resolution, thereby enhancing the reflected sensing signal. By contrast, in the beam-only optimization scheme, the sensing SINR continuously decreases because the sensing power remains fixed and cannot adapt to the increased system resources, while the growing communication power further suppresses the sensing performance.
\begin{table}[t]\centering
\footnotesize
\centering
\caption{Simulation Parameters}
\label{table:gesture_sim}
\begin{tabular}{|l|c|}
\hline
\textbf{Parameter}& \textbf{Value}\\
\hline  
$K$: number of users   & 4 \\
\hline
$M$: number of antennas  & 12 \\
\hline
$P_{\mathrm{max}}$: maximum Transmit Power   & 36 dBm \\
\hline
$\mathrm{SINR}_{\text{com}}^{\text{high}}$: high SINR threshold   & 5  \\
\hline
$\mathrm{SINR}_{\text{com}}^{\text{low}}$: low SINR threshold  & 1  \\
\hline
$\sigma_n^2$, $\sigma_r^2$: noise power    & $-90$ dBm \\
\hline
$T$: time slot duration   & 0.1 s \\
\hline
$K(f)$: absorption coefficient  & 0.02 \\
\hline
$h_A$: device height when picked up   & $1.5$ m \\
\hline
$h_B$: device height when put down & $1.2$ m \\
\hline
$\epsilon_h$: height variation threshold  & 0.1 m \\
\hline
\end{tabular}
\label{t1}
\end{table}
\begin{figure}[!t]
    \centering
    \includegraphics[width=0.5\textwidth]{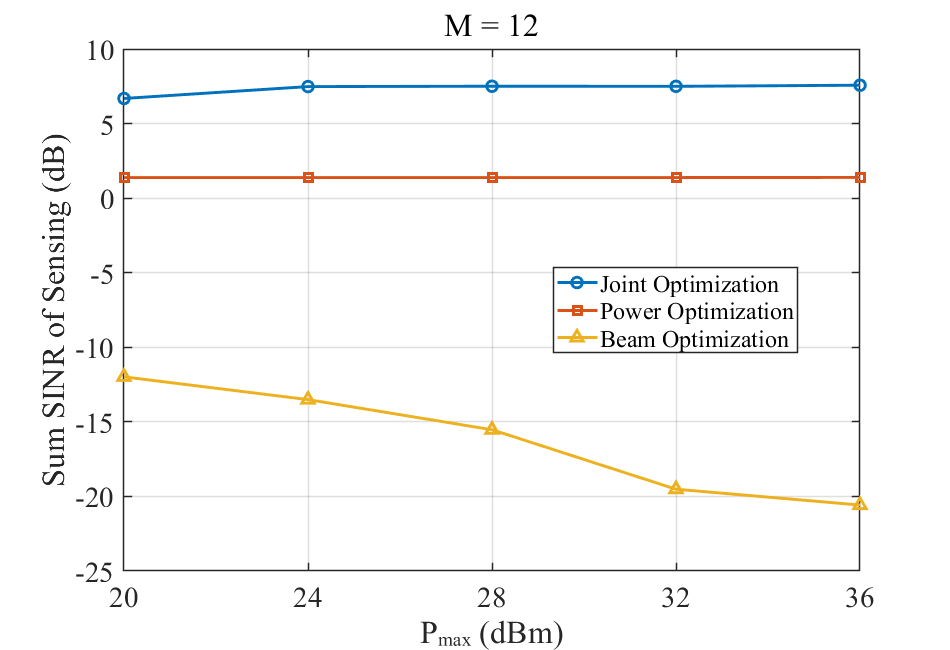}  % 替换为你自己的图片文件名
    \caption{Sum sensing SINR  versus $P_{\text{max}}$ }
    \label{fig2}
\end{figure}
\begin{figure}[!t]
    \centering
    \includegraphics[width=0.5\textwidth]{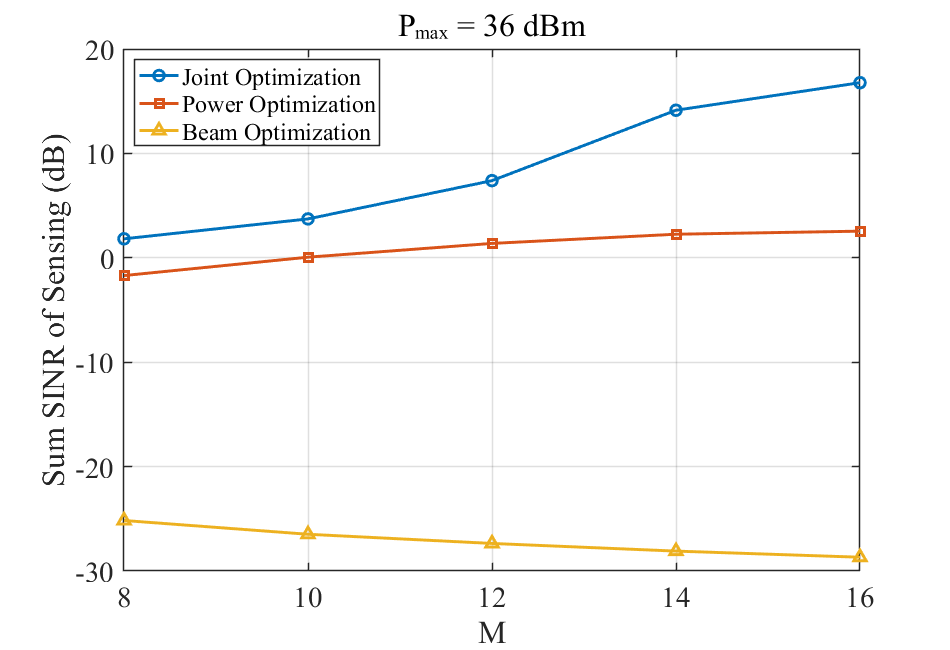}  % 替换为你自己的图片文件名
    \caption{Sum sensing SINR  versus $M$ }
    \label{fig3}
\end{figure}
\clearpage
\subsection{Performance under Dynamic Gesture Transitions}
We next consider a dynamic gesture scenario in which each user’s hand position varies across time slots. At timeslot 0, three users perform the action of putting down the device, while one user performs the action of picking up the device.

We conduct simulations under three different settings to evaluate the impact of antenna number and transmission power on system performance. As shown in Fig~\ref{fig4}, the sensing SINR remains low during the first three time slots, since the gesture variation does not exceed the predefined threshold and the system maintains a high communication rate. At time slot 4, the model successfully detects the gesture change and begins to adjust the power allocation and beamforming matrices. Compared with a static baseline without dynamic adaptation, the sum sensing SINR increases significantly and continues to improve in subsequent time slots. These results demonstrate the effectiveness and dynamic adaptability of the proposed model.

\addtolength{\topmargin}{0.10in}
\begin{figure}[!t]
    \centering
    \includegraphics[width=0.5\textwidth]{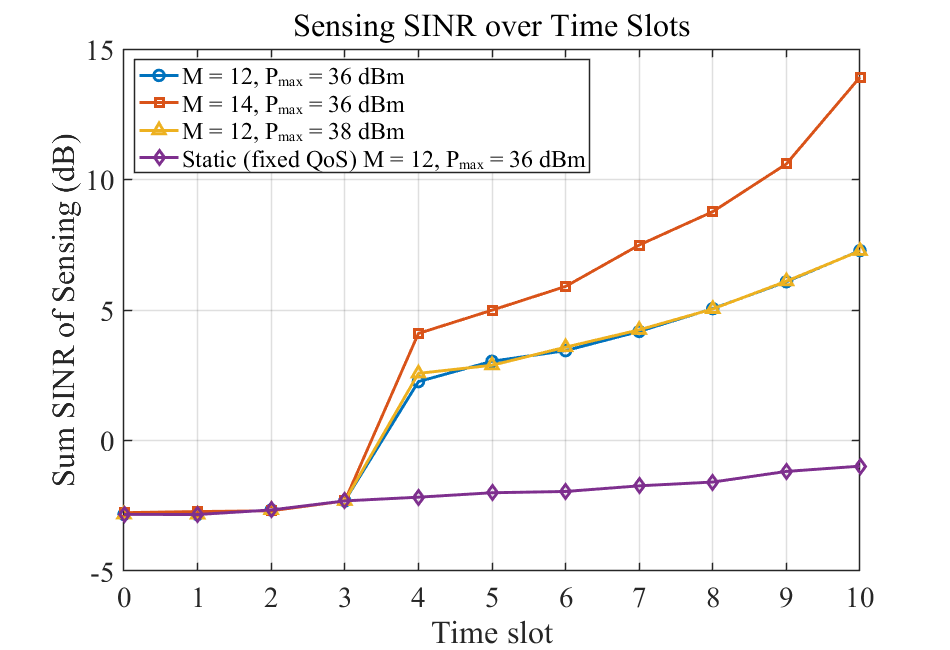}  % 替换为你自己的图片文件名
    \caption{Sum sensing SINR under Dynamic Gesture}
    \label{fig4}
\end{figure}
\section{conclusion and future work}
\enlargethispage{-0.50in}
This paper investigates a multi-user indoor THz-ISAC system for gesture recognition 
and proposes a joint optimization algorithm. 
The system adapts its strategy based on gesture changes, the proposed method improves sensing SINR while maintaining communication performance, and simulation results verify its effectiveness.  

The current model assumes fixed user height and stationary users with hand gestures only. 
In practice, variations in user posture and full-body motion introduce additional complexity. 
Future work will incorporate intelligent reflecting surfaces (IRS) and tracking mechanisms 
to improve robustness in dynamic environments.

\bibliographystyle{ieeetr}
\bibliography{main}

\end{document}